\documentclass{article}

\usepackage{PRIMEarxiv}
\usepackage[utf8]{inputenc} 
\usepackage[T1]{fontenc}    
\usepackage{hyperref}       
\usepackage{url}            
\usepackage{booktabs}       
\usepackage{nicefrac}       
\usepackage{microtype}      
\usepackage{rotating}
\usepackage{fancyhdr}       
\usepackage{graphicx}       
\graphicspath{{media/}}     
\usepackage{siunitx}
\usepackage{multirow}%
\usepackage{amsmath,amssymb,amsfonts}%
\usepackage{amsthm}%
\usepackage{mathrsfs}%
\usepackage[title]{appendix}%
\usepackage{xcolor}%
\usepackage{textcomp}%
\usepackage{manyfoot}%
\usepackage{algorithm}%
\usepackage{algorithmicx}%
\usepackage{algpseudocode}%
\usepackage{listings}%
\usepackage{subfig} 

\title{Stock Market Dynamics Through Deep Learning Context}

\author{
  Amirhossein Aminimehr\thanks{These authors contributed equally to this work.} \\
  School of Computer Engineering \\
  Iran University of Science and Technology \\
  Tehran, Iran\\
  \texttt{amir\_aminimehr@comp.iust.ac.ir} \\
   \And
  Amin Aminimehr\footnotemark[1] \\
  Department of Management \\
  Ershad Damavand Institute of Higher Education \\
  Tehran, Iran\\
  \texttt{aminaminimehr@outlook.com} \\
     \And
  Hamid Moradi Kamali \\
  School of Computer Engineering \\
  Iran University of Science and Technology \\
  Tehran, Iran\\
  \texttt{ha\_moradi@comp.iust.ac.ir} \\
     \And
  Sauleh Eetemadi \\
  School of Computer Engineering \\
  Iran University of Science and Technology \\
  Tehran, Iran\\
  \texttt{sauleh@iust.ac.ir} \\
    \And
  Saeid Hoseinzade \\
  Department of Finance \\
  Suffolk University \\
  Massachusetts, USA\\
  \texttt{shoseinzade@suffolk.edu} \\
  \\
}

\begin{document}
\maketitle

\begin{abstract}
Studies conducted on financial market prediction lack a comprehensive feature set that can carry a broad range of contributing factors; therefore, leading to imprecise results. Furthermore, while cooperating with the most recent innovations in explainable AI, studies have not provided an illustrative summary of market-driving factors using this powerful tool. Therefore, in this study, we propose a novel feature matrix that holds a broad range of features including Twitter content and market historical data to perform a binary classification task of one step ahead prediction. The utilization of our proposed feature matrix not only leads to improved prediction accuracy when compared to existing feature representations, but also its combination with explainable AI allows us to introduce a fresh analysis approach regarding the importance of the market-driving factors included. Thanks to the Lime interpretation technique, our interpretation study shows that the volume of tweets is the most important factor included in our feature matrix that drives the market's movements.
\end{abstract}


\section{Introduction}
The literature on financial market prediction using sentiment analysis is deeply rooted in the study of Schiller \cite{10.2307/1802789} and Merton\cite{10.2307/1913811}. They witnessed unstable price movements after the release of new information in the market. To speculate this notion, Schiller’s study questioned the assumption of rationality in the behavior of investors and the process of measuring present values of expected return in the efficient market model. Therefore, the firmly established belief of investors’ rational behavior was about to be challenged or disproven. Shiller’s study on the bond market and Stephen LeRoy's study \cite{10.2307/1911512} on the stock market quantitatively provided the primary evidences of the presence of new factors other than financial statements that are driving the price movements. It is worth noting, that schiller, in his famous research, stated that this new factor which is causing the extreme deviation from a rational expected return cannot be associated with data errors, price index problems, and changes in tax laws. 

In this respect, in 2004, Andrew Lo, by analyzing several researches that studied the psychological factors affecting financial and economic interactions, proposed a new idea called Adaptive Market Hypothesis(AMH). The prominent aspect of this idea is that investors do not always necessarily make decisions based on rational measurements, but their decisions are influenced by a combination of feelings \cite{Lo15}. Thus, financial and economic scholars required new factors to enhance the performance of their equilibriums. As a result, today, a crucial area of data science called sentiment analysis, as one out of many proxies for irrational decision analysis \cite{https://doi.org/10.1111/deci.12229}, has found numerous applications in the literature of finance \cite{articleTMF} and specifically market prediction.

On the other hand, Deep learning(DL) has led the scientific community to many breakthroughs over the recent years. Financial market analysis is one of the areas that has greatly benefited from this technology \cite{heaton_2016_deep}. DL approaches are widely used in financial time series prediction to model price, return, volatility, and risk, each of which taking the advantage of a specific variation of price data \cite{man_2019_financial}. DL models have often stood out for their extraordinary accuracy, exceeding conventional econometric methods\cite{sezer_2020_financial}. Alongside this, their strengths include the capability to incorporate unstructured text data into predictions, and a lessened susceptibility to issues associated with increasing variables, such as losing degree of freedom.

Coincidentally, as mentioned above in the literature of AMH and financial market prediction a need was felt to improve the quality of financial time-series forecasting models by incorporating the true sentiment of investors. Therefore, as one approach, analyzing social media content (through the use of Deep Learning) has been one of the attractive areas of the applications of DL in finance. Amongst many social media platforms that can be utilized as a source of investors' sentiment, Twitter is often favored due to its real-time content, ease of data retrieval, and abundance of rich financial tweets on it. Many studies have incorporated representations of market sentiment into their deep-learning models. These models can learn the non-linear underlying relations between market price and the sentiment of investors to incorporate them in their predictions. 

DL models have often proven to be more accurate than other well-known approaches such as classical econometrics approaches or machine learning methods \cite{RePEc:oup:rfinst:v:33:y:2020:i:5:p:2223-2273.}. Yet there seem to be still more capabilities to be exploited using them. Forecasting methods using machine learning can be better exploited because of the following two drawbacks: First, due to the abundance of the contributing factors on the market, these factors were not efficiently fed to the model which has led to insufficient accuracy. Second, DL models used in financial market analysis have been suffering from a lack of interpretation both in the learning process and prediction results, Which has been causing distrust amongst the beneficiaries of such models, such as financial market analysts and traders. \cite{ohana_2021_explainable} This fact is even more emphasized in practice since the applications of DL in finance directly deal with wealth and can result in significant financial loss. Therefore, as exceptional as deep learning models in comparison with classic econometrics methods, they are still not accurate enough. And, this lack of explanations of why and how DL methods produce their results contradicts the trustability that is emphasized in financial forecasting models. As a result of these challenges, there is still an aversion to the widespread utilization of such models in practice. However, there seem to be approaches to mitigate the mentioned issues which will be discussed below. 

Attempts to interpret deep learning models have opened the door to innovative solutions for the challenges of trustability and explainability in black box DL models. These solutions have resulted in more trustable and explainable applications of deep learning in financial time-series forecasting\cite{luo_2018_beyond, shi_2019_deepclue,basiri_2021_abcdm}, yet there have been fresh innovations in this area. These new approaches that utilize various explainable artificial intelligence (XAI) methods have not only shown promising results in gaining the trust of the end users, through their interpretation capabilities, but they have also caused the model to, directly or indirectly, enhance their accuracy compared to their non-interpretable counterparts. These studies have paved the way for more interpretable prediction approaches in the field that can help us build more trusted and accurate models.

In terms of financial predictions based on Twitter data, there seems to be a lack of explanation on what features in the Twitter data are more useful in predicting the behavior of financial markets or how and when these effects are in order. As such, features that include content interactions (Likes, Saves, comments, etc.), as well as other features such as the post's sentiment, the reputability of the content creator etc. are missing in many studies. In addition, the fact that many researchers have utilized their prediction methods with either market data (OHLC) or Twitter features separately has confined their interpretations to a limited scale. As a result, with all the mentioned advances in the field, there still seems to be a lack of a comprehensive study on the effects of various Twitter features on the results of financial DL models. 

Our novelty in this paper is introducing a feature matrix that is capable to mitigate the mentioned two limitations by the following upgrades:
\begin{itemize}
 \item First, this feature matrix is capable of incorporating both Twitter content data including sentiments of tweets, volume of tweets, number of likes, etc. as well as the features related to OHCL of market data. Thus, our model will be covering not only market factors, but also indices of the sentiment of investors and features related to tweets. This feature matrix also possess the capability to be further extended whenever necessary, in order to broaden the analyzed factors on the market.

 \item Second, this specific feature matrix provides us with the ability to implement an interpretation method that can comprehensively measure the contribution of each factor to the result. In this regard, our comparative study also supports the increase in the predictive power of our proposed feature matrix.

 \item Finally, as an important concern in financial market analysis, our empirical study through the advantage of the recently developed local-based interpretation method is capable of measuring the contributing factors on each instance of the prediction of the target value. Thus, our study will provide a feature analysis result that provides an insight into the dynamics of the studied samples of the stock market. 
\end{itemize}

The remaining of this study comprises the following sections. In section 2 we first summarize the studies on price prediction that have used text data analysis in their methodology. Then we review feature analysis methods and causality tests in the prediction studies along with their pros and cons. In section 3 we dive into our proposed methodology and its implementation. In section 4 we will analyze the results and compare them with our benchmark methods. In section 5 we will summarize the study.

\section{Related Works}
\label{sec:headings}
There is a rich literature on analyzing the effective factors in financial time series prediction. However, we believe such studies which include a broad range of variables have missed the sentiment factors and indices in their analysis \cite{RePEc:oup:rfinst:v:33:y:2020:i:5:p:2223-2273.} \cite{https://doi.org/10.1111/jofi.12883}. Therefore, what matters in our literature review are those studies that use sentiment analysis amongst their feature set in their predictions like the study by \cite{CHU2019101601} \cite{CHEN201814}.

In our literature review, we aim to earn an understanding of the methods used, markets analyzed, feature analysis methods used, and finally the results that have been obtained. The part of the results that mainly demonstrates the contributing factors to their studied financial market is a matter of focus for this study too. This will help us compare the result of our presented empirical study regarding the contributing factors with the factors mentioned in the literature. Furthermore, since we propose a novelty in the methodology of financial time series prediction that excels in the power of interpretability to a larger scale, we have also briefly mentioned the methodology, prediction and interpretation methods used in these studies.

The literature is divided into two main series of studies.

In the conventional methods of causality analysis like the Granger causality test by \cite{702ab909-8cb1-3c30-a5f1-ab4517d6cf1c} \cite{RePEc:spr:fininn:v:5:y:2019:i:1:d:10.1186_s40854-018-0119-8}  \cite{RePEc:spr:fininn:v:7:y:2021:i:1:d:10.1186_s40854-021-00275-9}, F-test is conducted to measure the significance of the difference of the explained variance of residuals of the model with and without the presence of the lags of a specific factor. If the performance of the fitness function is significantly improved by the presence of the lags of the factor in the model, compared to the one without that factor, we can make a statement. We state that the feature is significantly affecting the market's movement. This method is linearly analyzing the dependencies, but many studies have stated the fact that the financial market's dependencies are not linear and may vary through time. Thus, a universal dependency measure may not seem to be accurate and a comprehensive measurement approach. Furthermore, in cases that are solving a classification problem, the errors are discrete, and therefore not only the Granger causality test but also other non-parametric tests that are used for testing the significance of the difference of the errors are not applicable, like the Wilcoxon Sign rank test used in the method proposed by \cite{ROSOL2022106669}. The limited studies that include investors' sentiment indicators in their feature analysis and have used this causality approach are summarized in Table 1.

On the other side, studies that have incorporated Twitter content in their predictions, and at the same time have utilized DL methods like the study by \cite{article_from_soft_sentiment} are limited to just a prediction study and not feature interpretation in a nonlinear context. Moreover, it is worth noting that the only mentioned limited works that have attempted to explain or analyze the features through a DL framework have devoted themselves to analyzing the importance of words or sentences in tweets. In other words, their explanation methods measure the importance of each word in a sentence, or each sentence in a tweet on the sentiment polarity of the entire sentence or tweet respectively. In addition, there seems to be a lack of interpretation of general features used in Twitter textual and numerical data (number of likes, retweets, etc) in these studies. Nevertheless, we cover the mentioned literature in Table 2 although they may all not seem to resemble our point of view in their feature analysis method. But at least we can be sure that we have covered how deep learning and XAI experts have taken advantage of the modern feature analysis methods in financial prediction tasks.

We believe there is a lack of an interpretation framework that mitigates both the limitations of the methods that have been used thus far. Therefore, we have used the local interpretable model-agnostic explanations (Lime) method \cite{ribeiro_2016_why} to help us perform an interpretation study of a proposed custom feature matrix. Moreover, since the previous studies fall short of explaining why a tweet has been impactful on the market, our method explains the reason for the effectiveness of the tweet by interpreting the share of each feature on the result. This quality of Interpretability became feasible by providing a novel feature analysis based on the impact of features on the market’s movements.

\begin{sidewaystable}[htbp]
    \centering
    \caption{Related Works - Studies Incorporating Twitter Mood in Their Linear Predictions}
    \label{tab:academic_table}
        \begin{tabular}{p{0.5cm}p{0.5cm}p{3cm}p{2cm}p{1.5cm}p{3cm}p{2cm}p{1.5cm}p{5cm}}
        \toprule
        Study & Year & Description & Market & Analysis & Variables & Duration & Forecasting & Result \\
        \midrule
        \raggedright \cite{Gilbert2010WidespreadWA} & 2010 & \fontsize{8}{12}\selectfont Analyzing the linear correlation of the public mood inferred from LiveJournal posts with the market data and vice versa. & \raggedright Stock market – S\&P 500 index & Granger causality test & \raggedright A custom Anxiety Index ( A measure of aggregate anxiety and worry) & \raggedright 25 Jan – 13 June, 1 Aug – 30 Sep, 3 Nov – 18 Dec (2008) & - & {\fontsize{8}{12}\selectfont Their work which includes the 2008 global crisis shows that one standard deviation increase in the values of anxiety index can lead to 0.4\% lower returns in their target market. In addition, their work showed that there is no significant reverse causation from the market that induces a subsequent increase in the measure of anxiety. } \\
        
        \raggedright \cite{bollen_2011_twitter} & 2011 & \fontsize{8}{12}\selectfont Analyzing the linear correlation of DIJA index with the 7 mood indices captured from related tweets. & \raggedright Stock market – DIJA index & Bivariate Granger causality test & \raggedright 7 mood indices (1 index from OpinionFinder \cite{inproceedings} + 6 indices GPOMS) & \raggedright 28 Feb 2008 – 19 Dec (2008) & Fuzzy Neural Network & {\fontsize{8}{12}\selectfont  One of the 6 indices of GPOMS labeled as “Calm” showed significant causative relationship with the target value (DIJA index) in 2 to 5 days later.}\\
        
        \raggedright \cite{10.1371/journal.pone.0138441} & 2015 & \fontsize{8}{12}\selectfont Analyzing the linear correlation of 30 tickers from DIJA index with sentiment polarity of Twitter mood during specific events (Event study). & \raggedright Stock market – 30 stocks of the DJIA index & Pearson correlation, Granger causality & \raggedright Sentiment polarity of Twitter posts related to each ticker of DIJA index & \raggedright 1 June 2013 – 18 Sep 2014 & - & {\fontsize{8}{12}\selectfont  During the peak of the volume of Twitter posts, there seems to be a dependence between the sentiment polarity of the posts and the abnormal return of the ticker. This is evident during both, expected and unexpected increases in the volume of related tweets.} \\
        
        \raggedright \cite{pagolu2016sentiment} & 2016 & \fontsize{8}{12}\selectfont Analyzing the linear and non-linear correlation of tweet sentiments up to 3 days earlier related to Microsoft with the rise or fall of the stock of the same company. & \raggedright Stock market – Microsoft (MSFT) & LibSVM / Logistic Regression & \raggedright Sentiment of tweets (N-gram and Word2vec representations) & \raggedright 31 Aug 2015 – 25 Aug 2016 & LibSVM/ Logistic Regression & {\fontsize{8}{12}\selectfont  Their work showed that there is a strong correlation between the opinion of the related tweets about the company and the rise or fall of the price of the company in the stock market according to the model with relatively higher accuracy amongst their compared models.} \\
        
         \raggedright \cite{tabari-etal-2018-causality} & 2018 & \fontsize{8}{12}\selectfont Analyzing the linear correlation of two sentiment features on multiple stock returns and vice versa. &\raggedright Stock market – About 90 stocks & Granger causality test & \raggedright Sentiment of tweets labeled using Amazon Mechanical Turk and a custom built classification model & \raggedright 1 Jan 2017 – 31 Mar 2017 & - & {\fontsize{8}{12}\selectfont This study found significant causation relation between the sentiment of tweets and the events in the stock market, especially during the jumps in the market. This study also showed that reverse causation relation also exists from the market on the mood of the sentiments in the related tweets. } \\
    \end{tabular}
    \hspace*{0cm}\parbox{21cm}{\small Note: Study refers to Name of the Study, Year refers to Paper Date, Description refers to Brief Description of the Study, Market refers to Target Market, Analysis refers to Correlation/Causality Analysis Method, Duration refers to Duration of the Study, Forecasting refers to Method of Forecasting, and Result refers to Brief Result.}
\end{sidewaystable}

\begin{sidewaystable}[htbp]
    \centering
    \ContinuedFloat
    \caption[]{Related Works - Studies Incorporating Twitter Mood in Their Linear Predictions} 
    \label{tab:academic_table}
        \begin{tabular}{p{0.5cm}p{0.5cm}p{3cm}p{2cm}p{1.5cm}p{2cm}p{2cm}p{1.5cm}p{5cm}}
        \toprule
        Study & Year & Description & Market & Analysis & Variables & Duration & Forecasting & Result \\
        \midrule
        \raggedright {\fontsize{9}{12}\selectfont \cite{abraham_2018_cryptocurrency} } & 2018 & \fontsize{8}{12}\selectfont Analyzing the linear correlation and predictive power of three main sentiment proxies on the two largest coins in the Cryptocurrency market. & \raggedright Cryptocurrency market – Bitcoin and Ethereum & Linear regression & \raggedright Google trends, Sentiment of tweets (VADER score) and tweet Volume & \raggedright 60 consecutive days & Linear regression & {\fontsize{8}{12}\selectfont Their work showed that sentiment of tweets is not a reliable factor of explanation for the return data of the studied coins because they are biased towards a positive side while the volume of tweets and Google trends data seem to have predictive power for the target variable.}\\
        
        \raggedright {\fontsize{9}{12}\selectfont \cite{Kraaijeveld_2020_the} } & 2020 & \fontsize{8}{12}\selectfont Analyzing the linear correlation of Twitter post sentiments with the return of the cryptocurrency market. & \raggedright Cryptocurrency market – 9 largest coins & Granger causality test & \raggedright Volume, and Sentiment of tweets, VADER, and Custom index of bullishness & \raggedright 4 June 2018 – 4 Aug 2018 & - & {\fontsize{8}{12}\selectfont Their work showed that there is a linear causality relationship between some of the independent variables and some of the target variables in this study. There also seems to be a reverse relationship which means that the return of some coins in the market has caused changes in the volume of the related tweets. } \\
        
        \raggedright {\fontsize{9}{12}\selectfont \cite{DEOLIVEIRACAROSIA2021115470} } & 2021 & \fontsize{8}{12}\selectfont Analyzing the linear correlation of news sentiments with the Brazilian stock market and proposing investment strategies in this market using Sentiment Analysis. & \raggedright Stock market – Brazilian stock market (Ibovespa Index) & Granger causality test & \raggedright Index crafted using ANNs based on the sentiment of news in Brazilian news sources. & \raggedright Jun 2018 – Jun 2019 & - & {\fontsize{8}{12}\selectfont There is a linear relationship between the sentiment of the news of each day and the closing price of the same day and the opening price of the next day. There is also a mutual correlation showing that both sentiment of news and the price have double sided causality relationship.} \\
        
        \raggedright {\fontsize{9}{12}\selectfont \cite{unknown12556h1n1} } & 2022 & \fontsize{8}{12}\selectfont Analyzing the linear correlation of tweet sentiments with the movements of different indices in the stock market during H1N1 and COVID-19 pandemic. & \raggedright Stock market – IPC, NASDAQ 100, Down Jones, S\&P 500, etc. & Linear correlation & \raggedright Lexicon based indices & \raggedright Jun – July (2019), Jan – May (2020) & - & {\fontsize{8}{12}\selectfont The study finds that SenticNet correlates most with financial indices, and this correlation can be improved by shifting the series. It also confirms that a Twitter post's impact on a financial index is influenced by the number of followers, regardless of the post's accuracy. } \\
    \end{tabular}
    \hspace*{0cm}\parbox{21cm}{\small Note: Study refers to Name of the Study, Year refers to Paper Date, Description refers to Brief Description of the Study, Market refers to Target Market, Analysis refers to Correlation/Causality Analysis Method, Duration refers to Duration of the Study, Forecasting refers to Method of Forecasting, and Result refers to Brief Result.}
\end{sidewaystable}

\begin{sidewaystable}[htbp]
    \centering
    \caption{Related Works - Studies with Interpretable/Explainable Deep Learning Methods}
    \label{tab:academic_table}
        \begin{tabular}{p{0.5cm}p{0.5cm}p{2.5cm}p{2cm}p{1.5cm}p{3cm}p{2cm}p{2cm}p{5cm}}
        \toprule
        Study & Year & Description & Market & Analysis & Variables & Duration & Forecasting & Result \\
        \midrule
        \raggedright {\fontsize{8}{12}\selectfont \cite{shi_2019_deepclue} } & 2018 & \raggedright \fontsize{8}{12}\selectfont Proposed a visual technique to interpret the polarity of the tweets or news for stocks. & \raggedright S\&P 500 (Stock Market) & \raggedright Interactive, Hierarchical Visualization Interface & \raggedright News Headlines + tweets + Financial Reports & \raggedright 2006 to 2015 & \raggedright \fontsize{8}{12}\selectfont Hierarchical Feature Extraction + CNN + LSTM & {\fontsize{8}{12}\selectfont Provides a visually interactive hierarchical price prediction of stocks based on interpreted sentiment polarity of textual data (tweets, news, and reports).}\\
        
        \raggedright {\fontsize{8}{12}\selectfont \cite{yang_2018_explainable} } & 2018 & \raggedright \fontsize{8}{12}\selectfont Introduces a reduction in the noisy news for each day. And provides a more interpretable model. & \raggedright  S\&P 500 (Stock Market) & Attention Mechanism & \raggedright News of Reuters and Bloomberg in the financial sector &\raggedright 2006 - 2013 & \raggedright \fontsize{8}{12}\selectfont Dual-stage attention mechanism + GRU & {\fontsize{8}{12}\selectfont  Used attention layer to interpret the relevance of financial news and allocating weights to the different days in terms of their contribution to stock price movement. } \\

        \raggedright {\fontsize{8}{12}\selectfont \cite{10.5555/3370272.3370290} } & 2019 & \raggedright \fontsize{8}{12}\selectfont Predicting expert commentary for brands based on financial times-series (VAR and POS) And explaining the influence of each variable with SHAP & Private Company Sales & Shapley Values & \raggedright VAR from the company’s ERP, experts commentaries + Point of Sales & 2016-2019 & \raggedright \fontsize{8}{12}\selectfont KNN, SVM, Random Forest, and XGBoost,  & {\fontsize{8}{12}\selectfont proposes an explanation model using SHAP values to identify influential features in new datasets for financial time series data of a private company.} \\

        \raggedright {\fontsize{8}{12}\selectfont \cite{10.1145/3411174.3411191} } & 2020 & \raggedright \fontsize{8}{12}\selectfont Utilized a Hierarchical neural network alongside an LSTM model on Technical indicators and relevant news headlines & Thailand stock market (SET index) & LRP (Layer-wise Relevance Propagation) & \raggedright market data with (73 technical indicators) and Thai economic news headlines. & 1st of January 2008 to 31st of December 2019 & \raggedright \fontsize{8}{12}\selectfont LSTM and hirerarchical neural network. & {\fontsize{8}{12}\selectfont The paper demonstrates the interpretability of the hierarchical approach using the integrated gradient attribution methods, which is aimed at explaining the relevance of textual data to price movement} \\
        
        \raggedright {\fontsize{8}{12}\selectfont \cite{luo_2018_beyond} } & 2021 & \raggedright \fontsize{8}{12}\selectfont Provides a query-driven attention mechanism to interpret and visualize the results based on the given query by the end user & Chinese Stock Market & Query-based Attention Mechanism and text highlighting for visual representation & \raggedright Chinese mainstream financial websites (Financial Documents). & May 26 to June 25, 2017, & \raggedright \fontsize{8}{12}\selectfont GRU as both word level and sentence level sequence encoder. & {\fontsize{8}{12}\selectfont Proposed the FISHAQ} model, which takes the relevance of financial reports to the queries of the user as the subject of its visual method of interpretation.\\

    \end{tabular}
    \hspace*{0cm}\parbox{21cm}{\small Note: Study refers to Name of the Study, Year refers to Paper Date, Description refers to Brief Description of the Study, Market refers to Target Market, Analysis refers to Correlation/Causality Analysis Method, Duration refers to Duration of the Study, Forecasting refers to Method of Forecasting, and Result refers to Brief Result.}
\end{sidewaystable}

\begin{sidewaystable}[htbp]
    \centering
    \ContinuedFloat
    \caption[]{Related Works - Studies with Interpretable/Explainable Deep Learning Methods} 
    \label{tab:academic_table}
        \begin{tabular}{p{0.5cm}p{0.5cm}p{3.5cm}p{1.5cm}p{1.5cm}p{2cm}p{2cm}p{2.5cm}p{4cm}}
        \toprule
        Study & Year & Description & Market & Analysis & Variables & Duration & Forecasting & Result \\
        \midrule
        \raggedright {\fontsize{8}{12}\selectfont \cite{basiri_2021_abcdm} } & 2021 & \raggedright \fontsize{8}{12}\selectfont Proposed an Attention-based Bidirectional CNN-RNN Deep Model (ABCDM)(Utilizing both GRU and LSTM).& Customer long reviews & Attention Mechanism (to provide both forward and backward context) & \raggedright Long Reviews + tweets & - & \raggedright \fontsize{8}{12}\selectfont Attention-based bidirectional Convolutional Neural Network - Recurrent Neural Network (CNN-RNN) deep model & {\fontsize{8}{12}\selectfont Provides an interpretable (attention-based) method to detect sentiment polarity based on the textual content of both short tweets and long reviews.}\\
        
        \raggedright {\fontsize{8}{12}\selectfont \cite{ohana_2021_explainable} } & 2021 & \raggedright \fontsize{8}{12}\selectfont Used a GBDT model to forecast S\&P 500 stocks using 150 technical, fundamental, and macroeconomic features, and analyzed feature impact with SHAP. & S\&P 500 (Stock Market) & SHAP Values & \raggedright 150 technical, fundamental and macroeconomic features & detailed analysis of the March 2020 ﬁnancial meltdown & \raggedright \fontsize{8}{12}\selectfont gradient boosting decision trees (GBDT) & {\fontsize{8}{12}\selectfont Shows that retaining fewer and carefully selected features provides improvements across all Machine Learning (ML) approaches. consequently explaining the importance of each used feature in the study.} \\

        \raggedright {\fontsize{8}{12}\selectfont \cite{Gite2021StockPP} } & 2021 & \raggedright \fontsize{8}{12}\selectfont Aggregating news sentiment with historical indicators show better results than the technical indicators alone & National Stock Exchange & LIME & \raggedright {\fontsize{8}{12}\selectfont Aggregated 210,000+ Indian news headlines from Business Standard, The Hindu Business, Reuters, etc.} & - & \raggedright \fontsize{8}{12}\selectfont LSTM & {\fontsize{8}{12}\selectfont  Utilized an LSTM model on historical data and technical indicators alongside sentiment derived from financial news to predict the price and interpret the influence of each feature on the result using LIME.} \\

        \raggedright {\fontsize{8}{12}\selectfont \cite{RePEc:gam:jdataj:v:7:y:2022:i:11:p:160-:d:972099} } & 2022 & \raggedright \fontsize{8}{12}\selectfont Provides interpretation on the influence of each variable on the prediction results & Vietnam stock market & SHAP Values & \raggedright financial statements of Vietnamese Companies. & 2010 to 2021 & \raggedright \fontsize{8}{12}\selectfont Extreme Gradient Boosting + Random Forest & {\fontsize{8}{12}\selectfont Predicted financial distress for companies and explain the contribution of each feature to the result. long-term debts to equity, enterprise value to revenues, account payable to equity, and diluted EPS had greatly influenced the outputs } \\
        
    \end{tabular}
    \hspace*{0cm}\parbox{21cm}{\small Note: Study refers to Name of the Study, Year refers to Paper Date, Description refers to Brief Description of the Study, Market refers to Target Market, Analysis refers to Correlation/Causality Analysis Method, Duration refers to Duration of the Study, Forecasting refers to Method of Forecasting, and Result refers to Brief Result.}
\end{sidewaystable}

\section{Methodology}
In this work, we aim to propose three main novelties in the model implementation process, which will be discussed in this section.

As the first novelty of this study, we are generating a specific feature matrix capable of feeding our deep learning model with both historical price data and Twitter content. Regarding tweets, in addition to the sentiment extracted from the content itself, we are also incorporating the role of post interactions, post influences, and even the influence of the person who posts the tweet about the target tickers or company. In addition, we are also incorporating the historical return of price data up to three days earlier, so that we can extract any historical dependencies and patterns if they exist. This method of model feeding has many benefits. Firstly, since our method is in the category of deep learning models, its prediction performance is prone to increase by feeding it with more relevant data. Secondly, this paper aims to implement an XAI method to measure the influence of every factor on market data. So, we require a broad range of features from different sources that are introduced in the literature to see their effect using a single and unified measurement approach. Here in this paper, our measurement approach (XAI method) is Lime. Therefore, our feature matrix must be purposefully designed to cover the most renowned aspects of the literature of finance, so that the interpretation results can be illustrative, comprehensive, and in accordance with the expectations of scholars who are finance experts.

As the second novelty of this work, we implemented an explanation method on our deep learning model. This method can be used to explain the features that have led to a specific result and the importance of each feature in generating those results is measured. In this way, in addition to satisfying the concern of many scholars regarding what is causing a black box to generate a certain direction in the market, it also helps us know what is the most important factor that is influencing the market at a particular instance.

Thus, as the third novelty of this work, we provide an interpretation summary that illustrates the dependencies of the generated results on the input feature matrices in each instance separately. This feature is critical, specifically while interpreting the predictions on markets, since market movements are according to the sum of the investors' decisions, and the change in the sensitivity of investors' decisions builds a dynamic environment. As a result, an interpretation approach that is capable of showing these dynamic dependencies can not only interpret the causes of each generated result but it also can illustrate the changes in the dependencies through time and on each instance.

\subsection{Dataset, Variables and Feature matrix}
As part of the methodology and where most of the novelty of this study lies, we have provided the details of our data processing and the process of feature generation.
\subsubsection{Dataset and pre-process}
\textbf{Social Media Dataset:}
The social media dataset used in this paper has been downloaded from Kaggle.com and has also been part of the research project on factors influencing the market \cite{9378170}. From this dataset, Apple, Tesla, and Amazon, leading technology-sector companies, were selected due to their high volume of Twitter discussions. However, since the number of tweets in the first three years of the dataset regarding the mentioned tickers is deficient and therefore inefficient for prediction tasks, our research has only covered from June 1, 2018 up to December 31, 2019. In order to obtain tweets with higher impacts, first an index called total engagements was crafted based on the sum of the number of Likes, Comments, and Retweets of each post. Then, tweets with a total engagement of below 40 were omitted from the social media dataset. Next, the tweet texts were lowered cased and tokenized into separate words. Lastly, stop words and punctuation were removed to pass through the final steps which are stemming and lemmatization.

\textbf{Price Dataset:}
On the other side, the daily Open, High, Low, and Close price and trading volume of the same stocks were downloaded from Yahoo Finance without applying any justification with respect to dividends and splits. Then, the same duration as the duration of the mentioned Twitter dataset was cropped from the price dataset, to build a synchronous dataset of Twitter and market price. It is worth mentioning that the target variable in this study is the sign of the difference between Close price at time t minus Open Price of the same day, which turns this prediction problem into a binary classification problem. It has been shown by \cite{9165760} that while predicting a binary discrete variable rather than a continuous variable in financial time series using ML methods, the performance of the models is improved significantly.

\subsubsection{Variables and Feature matrix}
The variables used in our prediction method consist of 16 features that are either based on Twitter or market data. Our proposed feature matrix must be capable of being interpretable both time-wise and feature-wise. Therefore, features are distributed in the columns, while rows represent time. Our final feature matrix is a combination of two different feature matrices that are each specified for a specific data source.

Our first feature matrix consists of all information regarding Twitter content. In this method, inspired by the study of \cite{GAO202048} the intraday effects of Twitter post indices have been measured by feeding the models with Twitter post factors with 2-hours separated features within a day. In this respect, we measured our Twitter post indices within 2-hour time intervals, leading to a matrix of 12 time stamps(through the rows) for each feature(through the columns). To choose what factors from the Twitter post should be included in the feature matrix, the results from the studies mentioned in Table 1 were the basic guidelines. For instance, volume of tweets has been one of the main components of neural networks \cite{articleBVT} and its role has been mentioned in \cite{10.1371/journal.pone.0138441} and \cite{tabari-etal-2018-causality}  to be significant in affecting the market movements. Also, other studies such as \cite{Gilbert2010WidespreadWA},\cite{bollen_2011_twitter},\cite{pagolu2016sentiment},\cite{tabari-etal-2018-causality},\cite{Kraaijeveld_2020_the},\cite{DEOLIVEIRACAROSIA2021115470},\cite{unknown12556h1n1} have shown the significance of the impact of investors' sentiments on the market movements; therefore, we added this feature to our novel feature matrix. However, since this is an interpretation study, other factors that were not mentioned in the earlier literature were also incorporated in the model in the hope of a more accurate prediction and, thus, to make a more comprehensive analysis of the contributing factors of the studied market. As a result, apart from the raw factors of the tweets such as the number of likes, comments, re-tweets, and the volume of tweets, there is a custom index crafted based on the historical reactions of the community to the account that is posting each tweet. In other words, we crafted an efficacy index for each tweet that is measured by summing the historical engagement of the person who is posting a new tweet. It is worth mentioning that, while measuring this index we do not include the engagement of the last post itself and later posts interaction in the calculations to avoid any possible knowledge leakage. We named this score as writer score, which is calculated based on equations 1 and 2 respectively. In this formula let WS be the writer score, TE be the total engagement of each post, L be the number of likes, C be the number of comments and RT be the number of retweets of each post. Then we will have the TE calculated as shown in equation 1.

\begin{equation}
TE=L+C+RT
\end{equation}
After having the TE of each post measured separately, the writer score is calculated based on equation 2.

\begin{equation}
WS_t=\sum_{t=0}^{t-1}{TE_t}
\end{equation}

\begin{figure}[h]
  \includegraphics[width=\linewidth]{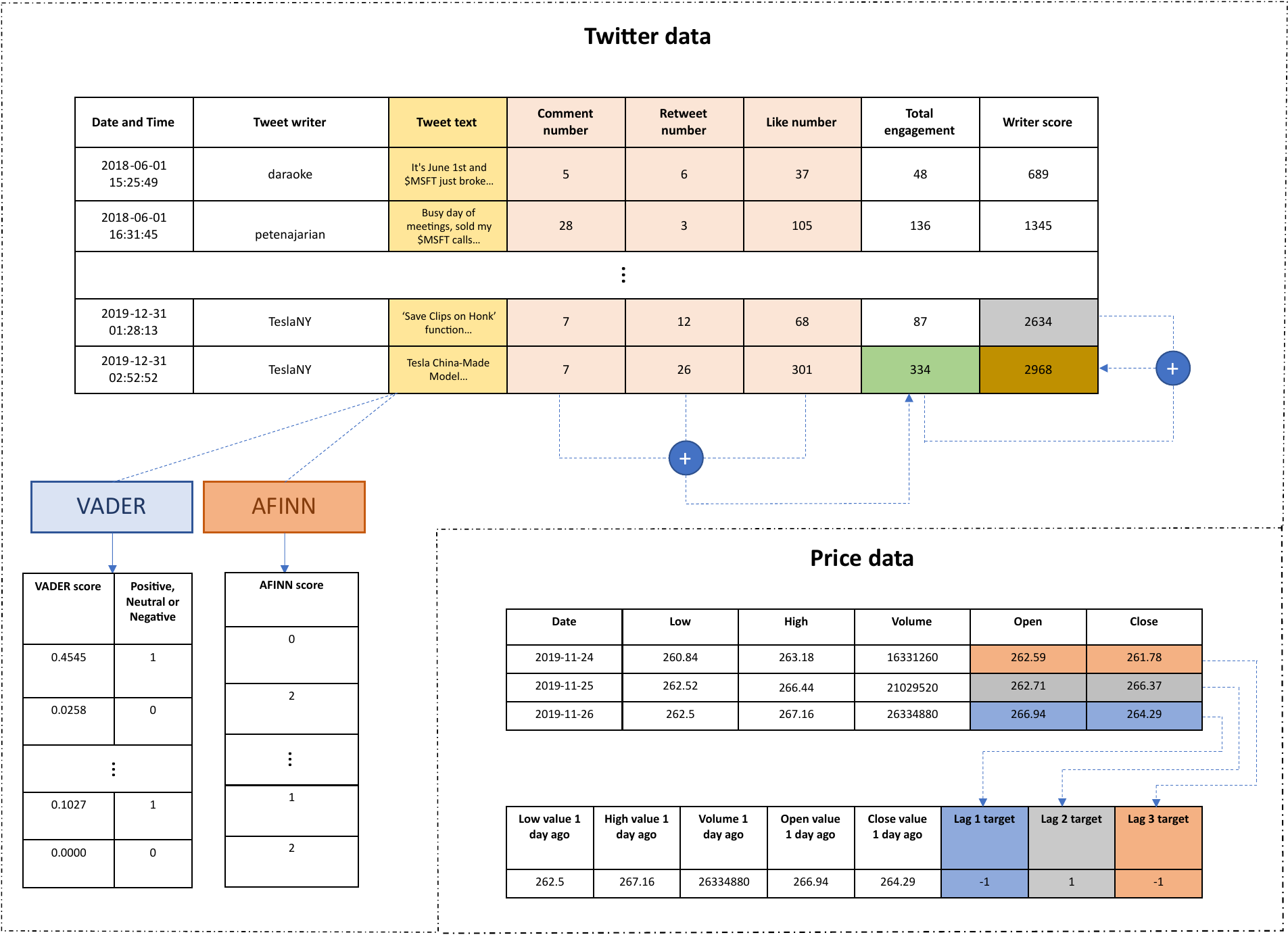}
  \caption{Variables used to build the feature matrix}
  \label{fig:boat1}
\end{figure}

This proposed feature matrix also constitutes three indices of tweet sentiments. The first index is the measure that corresponds to the aggregate of the VADER scores of all related tweets in each time interval of the measurement. The second index measures the Afinn score of the same tweets in the same time frame. Finally, as the third index, we first built a discrete dummy variable-based index for each tweet that is formed using VADER score and is the measure of the overall positivity or negativity of each tweet. Next, based on this score, 1 corresponds to an overall positive sentiment in the tweet, 0 corresponds to neutral, and -1 corresponds to a negative sentiment in the respect tweet. The final index in each 2-hour interval is measured by summing up this mentioned index in all the tweets during each 2-hour time interval of the day. Figure \ref{fig:boat1} illustrates the process of building the mentioned indices in detail on a briefly depicted table of the tweets dataset. In this approach, by implementing the interpretation method, we can observe if short-term intra-day Twitter features distinctly affect the market movements.
\newpage

At the same time, we formed a second feature matrix that constitutes 8 features, again through the columns. This feature matrix constitutes the lags of the target variable up to 3 days earlier so that the effect of memory can be taken into account. The rest of the features included in the second feature matrix are the OHCL prices of the prior day and the trading volume. In building this matrix with an intraday time frame, we had a limitation in access to hourly data, while on the other side, we were about to implement a CNN-based method. As a result, we synthesized this matrix by interpolating the same data 12 times throughout the day to match the first feature matrix dimension. 

\begin{figure}[h]
  \includegraphics[width=\linewidth]{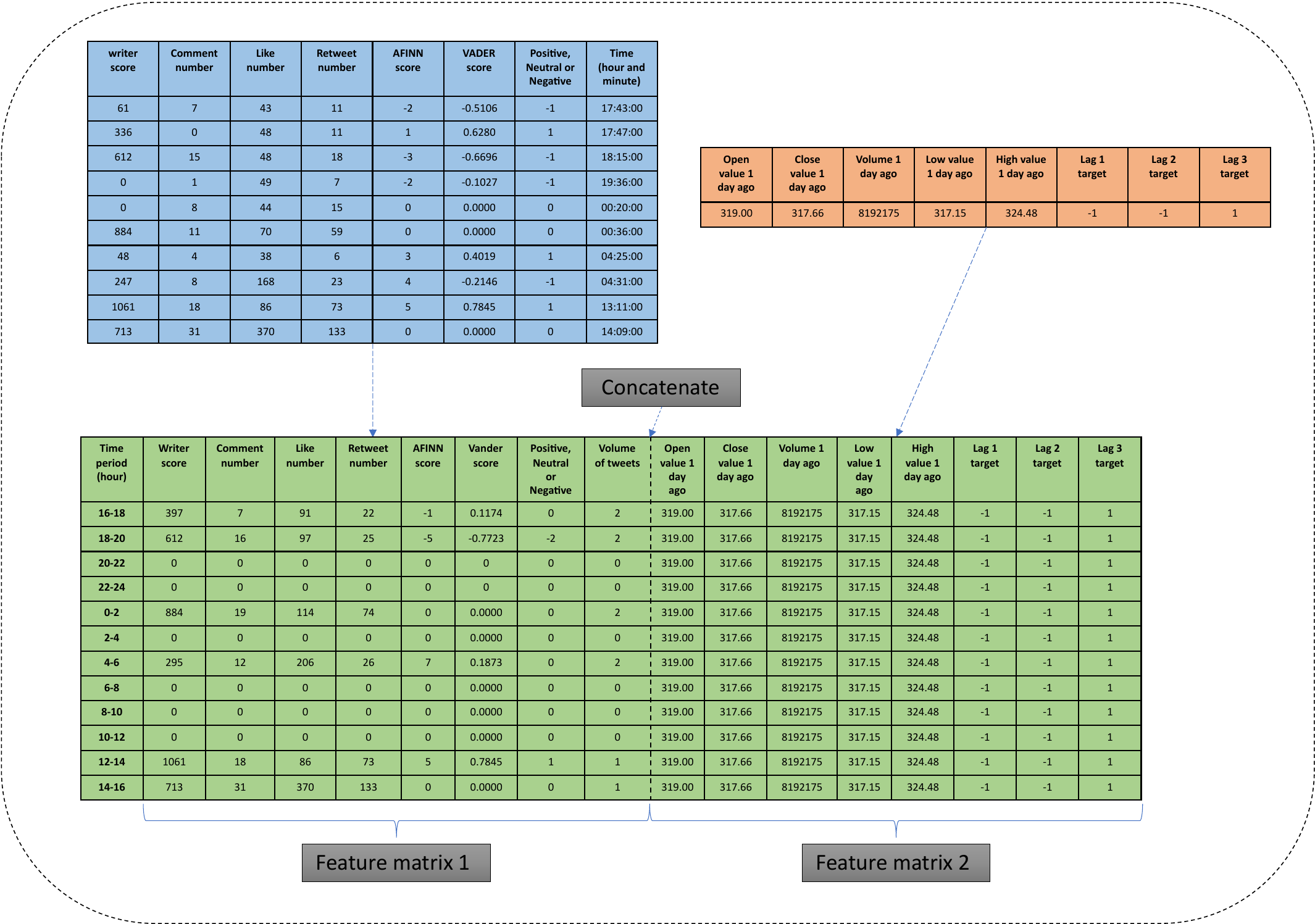}
  \caption{A sample of the final feature matrix at t-1}
  \label{fig:fig-3}
\end{figure}

We, finally, concatenated both the feature matrices along their time axis which had the same length of 12 to form the final 12 * 16 feature matrix. We call this feature matrix the time t-1 which will be mapped to the target value at time t. Figure \ref{fig:fig-3} shows the detail of each of the features that are used in the feature matrix. By combining the mentioned features in the form of a 2D feature matrix depicted in this figure our input matrix was crafted. This type of feature feeding enables us to use a deep learning method such as a CNN to extract the deep relationships between each feature across both the time and the feature dimensions using kernel calculations.

\newpage

\subsection{Prediction method}
\subsubsection{Main methods}
In our study on real world data, we first require a deep neural network architecture that can accurately fit the data on both train and test sets. Since perturb based methods provide their interpretation results based on the measure of the contribution of each factor after the model is trained, finding an accurate model is a compulsory step. Therefore, we first conduct a study using different deep neural network and then, using the most accurate model, we conducted our interpretation study. 

Two Deep learning models were implemented in this regard, a CNN-LSTM based architecture and a CNN based architecture. Based on the size of input feature matrices either one or two convolutional layers were incorporated. Each block also contained one pooling layer. Based on our feature matrix that consisted up to three lags of the target variable, the LSTM in the CNN-LSTM architecture in our proposed method contained three LSTM blocks. The feature matrix fed to each block of LSTM in the CNN-LSTM based architecture was embeddings extracted from feature matrices t-1, t-2 and t-3, respectively. Finally,  greed search was used to optimize the measure of regularization. 

\begin{figure}[h]
  \includegraphics[width=\linewidth]{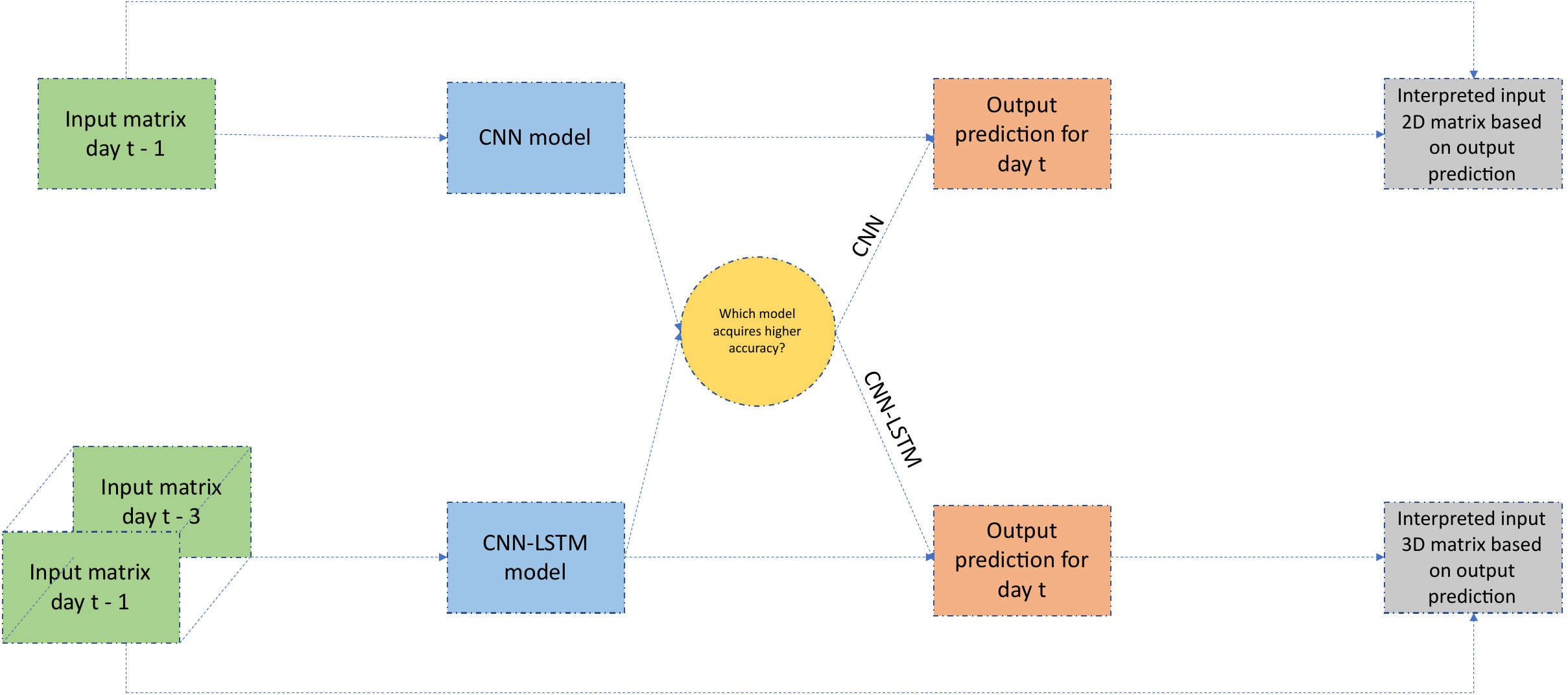}
  \caption{Flow chart of the process of prediction and feature interpretation}
  \label{fig:fig-1}
\end{figure}

\newpage
The flowchart of the prediction methods is illustrated in Figure \ref{fig:fig-1}.

\subsubsection{Benchmark methods}
In order to evaluate our feature matrix in extracting the knowledge of Twitter posts, we used other feature matrices based on \cite{10.1007/s11280-021-00880-9} \cite{article123451122} studies in place of the first feature matrix in our proposed method. Therefore, BOW and Doc2vec methods were used to extract embedings from the studied tweets. However, we limited the extracted embeddings to be in three different sizes of 8, 16 and 24 to be analogous with our second feature matrix. In this way, we were enabled to concatenate the extracted embeddings with the second matrix that stores the historical price data.
In addition to the mentioned benchmark methods, in order to check if there is any dependencies between stocks movements and the corresponding Twitter contents we also conducted an experiment without the presence of the first feature matrix. Finally, as the last experiment, we implemented the model with only sentiment of the tweets and the historical data feature matrix, to see if other feature in the Twitter posts such as volume of tweets, number of likes etc were relatively new in this context are effective in increasing the prediction accuracy or not?

\subsection{Interpretation method}

In this paper, we take advantage of one of the latest and most renowned interpretation methods, Lime(local interpretable model-agnostic explanations). Lime is a model developed by \cite{ribeiro_2016_why}, and it is among the local-based methods. Lime's methodology is based on a secondary human-interpretable model that is purposefully estimated on the local perturbations of the feature matrix. Lime can provide a local sensitivity analysis of the change in each feature to the change in the predicted probability of producing the target variables. As a result, Lime provides a measure of the contribution of each feature in each feature matrix in producing the target value. 

In our methodology we use Lime for the following objectives:

First, we need to analyze each feature's overall contribution, namely the number of likes, comments, VADER score, Afinn score, trading volume, lags of target, etc. In this way, we can understand that our model as the representative of the market movements is derived mostly by which of the factors.

Second, we need to analyze the changes in the intra-day effects of features related to tweet contents while they approach the end of the intra-day study duration.

Finally, we intend to analyze the behavior of the importance that each feature, used in our feature matrix, contributes to the final prediction with an inter-day perspective. In this approach, we will be able to illustrate the change in the dependencies of the market return on each target through the duration of test data.

Due to the relatively high number of the variables used in our model and the mentioned problem of losing degree of freedom, we can not include linear regression (standard coefficients) in our comparative study. However, the results of this empirical study will cover a comparative analysis on the contributing features that our pipeline has found and what the literature has witnessed. It is worthy to note that this method of interpretation has specific drawbacks which will be discussed in the final section as the limitations of this work.

\section{Results}
In this section, we analyze the accuracy of the models from a high-level point of view up to a detailed perspective while following the chronological order of the methodology.
\subsection{Prediction Results}

This empirical study analyzed three stocks, namely Apple, Tesla, and Amazon, as the samples. The performance of each model in the defined binary classification problem has been summarized in Table 3. In order to avoid the experimental errors in our DL method, inspired by \cite{9006342}, we trained each model ten times and finally reported the average of the generated results. In this Table, across the columns along with each of the sample tickers we provide the details of the incorporated DL architecture. In addition, across the rows of this Table, we walk through different feature matrices fed to our models. This enables us to better analyze the results with different perspectives.

Based on the accuracy of the predictions on the test set, firstly, it is evident that CNN-based architecture outperformed CNN-LSTM-based architecture while being exposed to our proposed feature matrix across all three studied tickers. The success of CNN in gaining the most accurate results in this experiment complied with the study of \cite{shi_2019_deepclue}. However, this is not necessarily witnessed when including the rest of the feature sets. Secondly, while comparing the results of the models that benefit from our proposed feature matrix and the model that has only the price data feature set, across all studied companies and DL architectures, we infer that Twitter data has extractable knowledge that can assist the deep learning models to enhance the estimation of the direction of our sample data, which aligns with the result of the study of \cite{article_from_soft_sentiment}. Likewise, by comparing the accuracy of the models that are fed with only embeddings of the text of the tweets (BOW8, BOW16, BOW24, DOC2VEC 8, DOC2VEC 16, DOC2VEC 24)  with our proposed feature matrix that instead of embeddings consisted indices such as the sentiments polarity of the text, number of tweets, likes and comments, and the historical influence of the tweet writer we infer that our feature matrix has more effective knowledge to be extracted, which has led to higher accuracy in predictions. The fact that which one of the features included in our proposed feature matrix that may be responsible for this increase in the accuracy of the predictions can be measured in our study using the interpretation step. 

\begin{table}[h]
\centering
\caption{Prediction results}
\label{tab:my-table}
\renewcommand{\arraystretch}{1} 
\begin{tabular}{@{}ccccccc@{}}
\toprule
\multirow{2}{*}{Test Set} & \multicolumn{2}{c}{Apple} & \multicolumn{2}{c}{Amazon} & \multicolumn{2}{c}{Tesla} \\ \cmidrule(lr){2-3} \cmidrule(lr){4-5} \cmidrule(lr){6-7}
& CNN & CNN-LSTM & CNN & CNN-LSTM & CNN & CNN-LSTM \\ \midrule
Proposed feature matrix & \textbf{61.19} & 60.731 & \textbf{52.926} & 52.439 & \textbf{60.714} & 59.512 \\
BOW 8 & 57.38 & 60.243 & 49.523 & 47.804 & 52.38 & 57.317 \\
BOW 16 & 58.095 & 58.292 & 48.809 & 51.951 & 58.571 & 57.56 \\
BOW 24 & 55.476 & 57.804 & 49.285 & 51.951 & 50.238 & 55.609 \\
DOC2VEC 8 & 55 & 58.536 & 50.714 & 49.512 & 54.523 & 50.243 \\
DOC2VEC 16 & 59.285 & 57.317 & 52.619 & 50.731 & 57.619 & 57.317 \\
DOC2VEC 24 & 58.571 & 56.097 & 50.238 & 47.317 & 57.619 & 55.365 \\
Sentiment + Price & 56.428 & 58.048 & 49.285 & 49.024 & 58.095 & 58.048 \\
Price only & 52.619 & 53.658 & 45.714 & 47.56 & 55.714 & 56.829 \\ \midrule
\multirow{2}{*}{Validation Set} & \multicolumn{2}{c}{Apple} & \multicolumn{2}{c}{Amazon} & \multicolumn{2}{c}{Tesla} \\ \cmidrule(lr){2-3} \cmidrule(lr){4-5} \cmidrule(lr){6-7}
& CNN & CNN-LSTM & CNN & CNN-LSTM & CNN & CNN-LSTM \\ \midrule
Proposed feature matrix & 57.837 & 61.621 & 58.378 & 56.486 & 62.432 & 60.54 \\
BOW 8 & 62.702 & 62.432 & 60.27 & 59.459 & 62.972 & 62.162 \\
BOW 16 & 56.216 & 57.297 & 59.189 & 65.135 & 60.54 & 61.621 \\
BOW 24 & 59.459 & 59.999 & 59.999 & 60.27 & 56.216 & 61.621 \\
DOC2VEC 8 & 59.729 & 61.891 & 60.27 & 58.108 & 59.189 & 64.594 \\
DOC2VEC 16 & 57.027 & 69.189 & 64.054 & 61.081 & 61.351 & 60.27 \\
DOC2VEC 24 & 59.459 & 56.486 & 62.972 & 61.621 & 59.189 & 64.864 \\
Sentiment + Price & 59.729 & 63.513 & 55.945 & 54.864 & 61.621 & 61.351 \\
Price only & 56.216 & 59.189 & 59.729 & 52.702 & 58.378 & 58.648 \\ \midrule
\multirow{2}{*}{Train set} & \multicolumn{2}{c}{Apple} & \multicolumn{2}{c}{Amazon} & \multicolumn{2}{c}{Tesla} \\ \cmidrule(lr){2-3} \cmidrule(lr){4-5} \cmidrule(lr){6-7}
& CNN & CNN-LSTM & CNN & CNN-LSTM & CNN & CNN-LSTM \\ \midrule
Proposed feature matrix & 51.381 & 61.265 & 67.837 & 69.006 & 51.981 & 56.174 \\
BOW 8 & 63.963 & 54.126 & 58.078 & 64.397 & 71.351 & 67.198 \\
BOW 16 & 55.135 & 58.734 & 56.636 & 70.090 & 64.864 & 69.518 \\
BOW 24 & 74.024 & 63.704 & 61.681 & 60.301 & 79.279 & 62.198  \\
DOC2VEC 8 &  55.675 & 58.674 & 56.546 & 59.518 & 60.150 & 69.939 \\
DOC2VEC 16 & 54.654 & 73.373 & 68.168 & 64.307 & 67.327 & 60.783  \\
DOC2VEC 24 & 66.276 & 52.349 & 52.912 & 51.475 & 65.885 & 65.361 \\
Sentiment + Price & 58.858 & 57.048 & 52.552 & 53.222 & 61.111 & 68.262 \\
Price only & 51.081 & 53.192 & 48.198 & 49.156 & 49.939 & 51.054 \\ \bottomrule
\end{tabular}
\end{table}
\clearpage

\subsection{Interpretation Results}
By selecting the CNN-based architecture, we step forward to the next step. Here, we require the predictions of the selected model to implement our interpretation method. The interpretation results, as mentioned earlier, are based on the Lime method. Since this method is capable of interpreting the mapping function locally, we feel free to interpret the results with both time-based and feature-based perspectives. 

Firstly, we provide a high-level interpretation of each feature which is the accumulation of the instances in which the predictions have truly predicted the target. In other words, we see the contributions of features by summing up the importance of the feature only in the predictions that have led to a true result, either TP or TN. The results of this overall contribution of the features in our novel feature matrix for each predicted stock is summarized in Table 4. As summarized in this Table, the contribution of the features seem to be close to each other; however, the importance of the volume of tweets seems to be distinctively more than the rest of the features of the feature matrix. Even though our implemented method is non-linear, this result seems to be congruous with the studies of \cite{10.1371/journal.pone.0138441} and \cite{tabari-etal-2018-causality} which had significantly shown the importance of Volume of tweets on the price movements of stock market, with a linear approach. 

Next, based on Table 5, we analyze the overall importance of the features but with a distinction made on the inter-day time of the tweets during the 24-hour time interval that the feature matrix covers. Although it is hard to make a clear distinction between the results, there seems to be a slight drop in average of the importance of the feature when the market time starts. We believe that this drop may be attributed to the fact that our feature matrix consist of the inter-day data of the tweets, but the price data has a daily time frame. Therefore, we continue this speculation by mentioning that tweets released during the market time may instantly affect the price and expire quickly; therefore, leaving no effect on the change in the daily scale of price data. However, this is not the case for the tweets that are released on after-hours. We finalize this speculation by mentioning that when a tweet is released in the after-hour period there is no active market to reflect it on the price unless the market opens the next day. Finally, the point that can be inferred from this Table is that the features in our feature matrix which are related to the time duration of 20-22 of the night prior to the next trading day are conspicuously more than the rest of the hours in all three studied samples.

\begin{figure}[h]
  \includegraphics[width=0.7\linewidth]{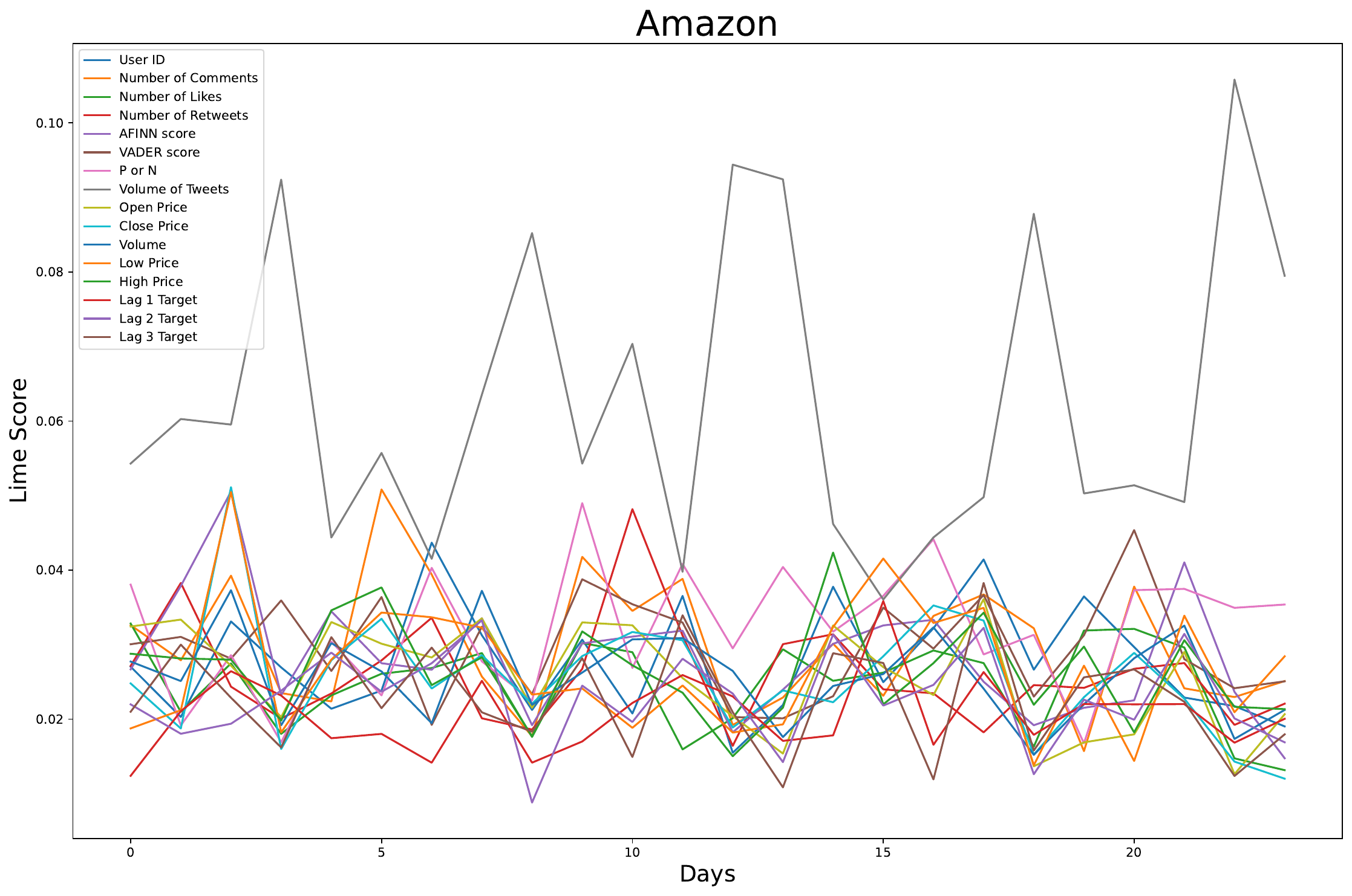}
  \centering
  \caption{Instance-based feature importance of Amazon through time (The line plot only illustrates the True predictions)}
  \label{fig:fig-4}
\end{figure}

\begin{figure}[h]
  \includegraphics[width=0.7\linewidth]{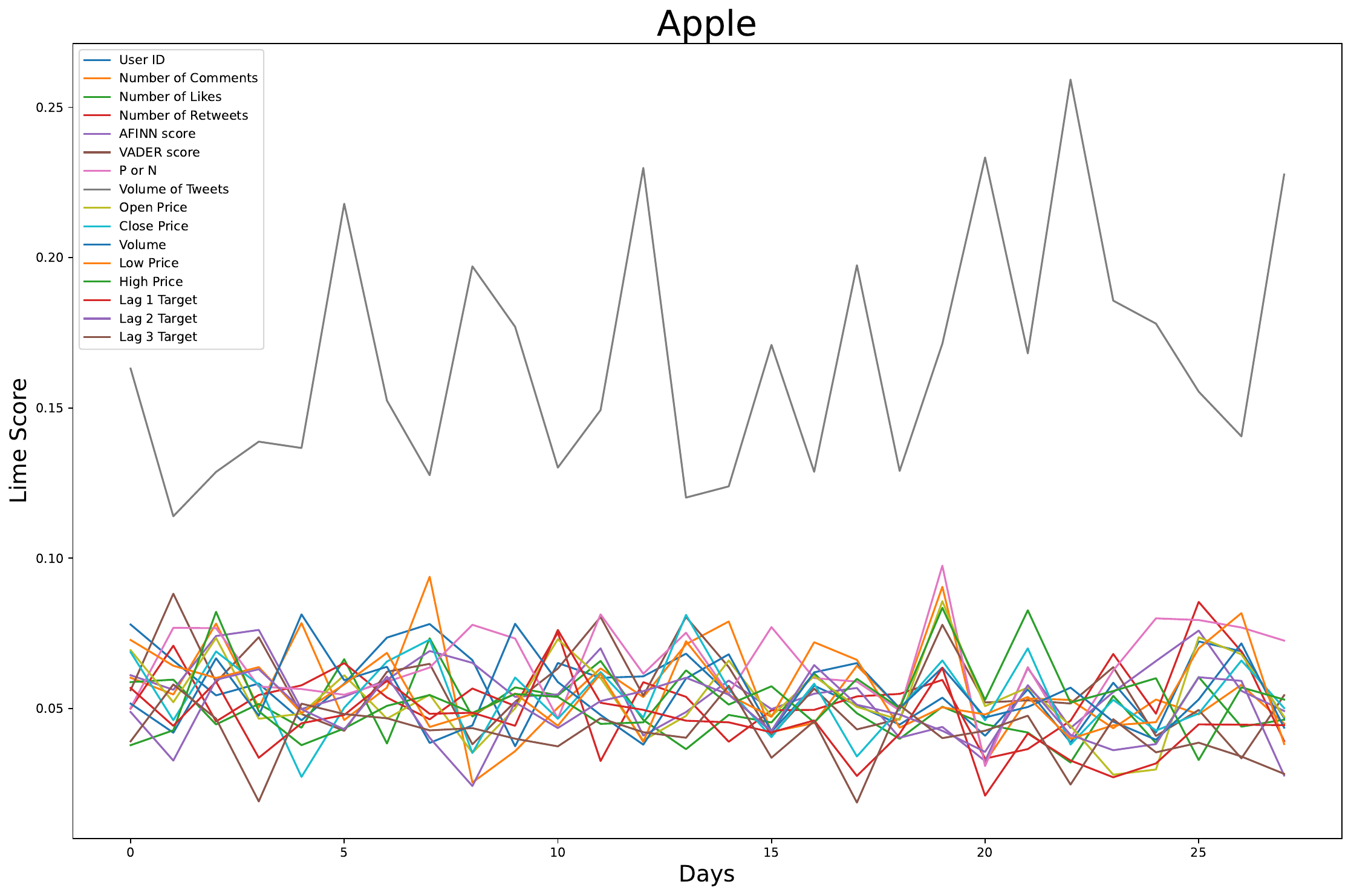}
  \centering
  \caption{Instance-based feature importance of Tesla through time (The line plot only illustrates the True predictions)}
  \label{fig:fig-5}
\end{figure}

\begin{figure}[ht]
  \includegraphics[width=0.7\linewidth]{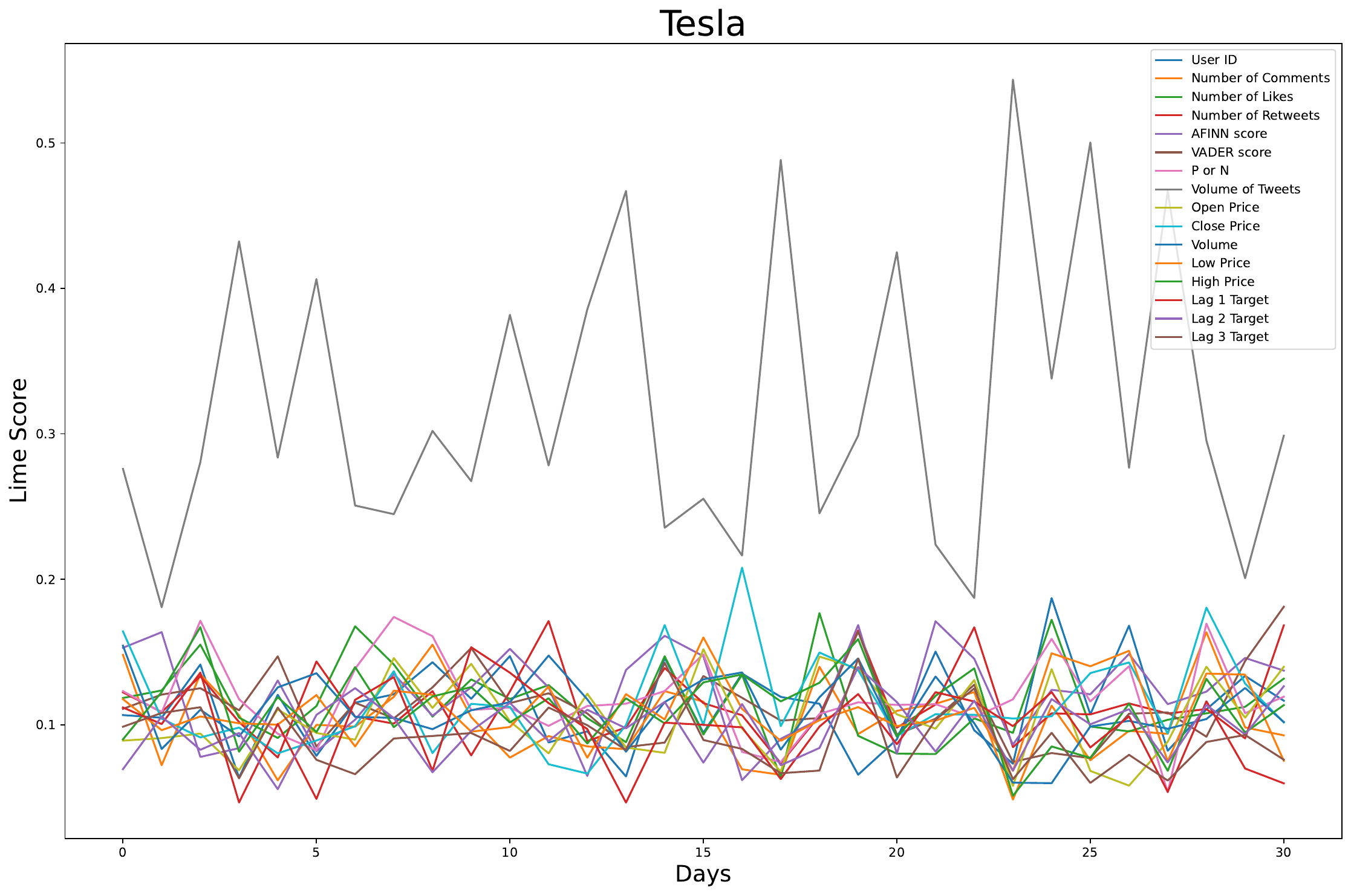}
  \centering
  \caption{Instance based feature importance of Apple through time(The line plot only illustrates the True predictions)}
  \label{fig:fig-6}
\end{figure}

Ultimately, by analyzing the figures \ref{fig:fig-4} to \ref{fig:fig-6} which provide an instance-based summary of the importance of features that has led to a successful prediction, we can see that the dynamics of the market seem not to change much through time. By this, we mean that through the instances of the test data, the models have not changed their reliance on the governing features considerably. This was inferred as the plot seemed to have fluctuated in a determined bond when considering each feature separately. In addition, as also mentioned earlier, the importance of all features seems to be pretty close to each other while the importance of the volume of tweets seems to be distinctively more than the rest of the factors in all the three studied firms.
\newpage

\clearpage

\begin{table}[h]
\centering
\renewcommand{\arraystretch}{1} 
\caption{Feature Importance (Feature-wise interpretation)}
\label{tab:my-table-2}
\begin{tabular}{@{}l S[table-format=1.4] S[table-format=1.4] S[table-format=1.4]@{}}
\toprule
Feature & {Amazon} & {Apple} & {Tesla} \\
\midrule
User ID & 0.0150 & 0.360 & 0.0836 \\
Number of comments & 0.0150 & 0.0376 & 0.0822 \\
Number of Likes & 0.0141 & 0.0352 & 0.0822 \\
Number of Re-Tweets & 0.0135 & 0.0342 & 0.0845 \\
AFINN Score & 0.0158 & 0.0405 & 0.0888 \\
VADER Score & 0.0143 & 0.0387 & 0.0838 \\
Positive or Negative polarity & 0.0166 & 0.0457 & 0.0945 \\
Volume of tweets & \textbf{0.0360} & \textbf{0.1108} & \textbf{0.2432} \\
Open price & 0.0140 & 0.0362 & 0.0880 \\
High price & 0.0150 & 0.0340 & 0.0765 \\
Low price & 0.0143 & 0.0361 & 0.0801 \\
Close price & 0.0140 & 0.0351 & 0.0840 \\
Volume of trading & 0.0139 & 0.0372 & 0.0814 \\
Lag 1 - Target & 0.0119 & 0.0301 & 0.0731 \\
Lag 2 - Target & 0.0136 & 0.0305 & 0.0732 \\
Lag 3 - Target & 0.0120 & 0.0314 & 0.0739 \\
\bottomrule
\end{tabular}
\end{table}
\clearpage

\clearpage
\begin{table}[h]
\centering
\renewcommand{\arraystretch}{1.2} 
\caption{Time Importance (Time-wise interpretation) }
\label{tab:my-table}
\begin{tabular}{@{}l S[table-format=1.4] S[table-format=1.4] S[table-format=1.4]@{}}
\toprule
Time Period & {Amazon} & {Apple} & {Tesla} \\
\midrule
16-18 (Market Closed) & 0.0209 & 0.0480 & 0.1117 \\
18-20 (Market Closed) & 0.0208 & 0.0459 & 0.1130 \\
20-22 (Market Closed) & \textbf{0.0399} & \textbf{0.0846} & \textbf{0.1931} \\
22-24 (Market Closed) & 0.0202 & 0.0693 & 0.1512 \\
0-2 (Market Closed) & 0.0197 & 0.0541 & 0.1310 \\
2-4 (Market Closed) & 0.0192 & 0.0542 & 0.1201 \\
4-6 (Market Closed) & 0.0208 & 0.0504 & 0.1123 \\
6-8 (Market Closed) & 0.0226 & 0.0511 & 0.1109 \\
\midrule
8-10 (Market Open) & 0.0213 & 0.0470 & 0.1094 \\
10-12 (Market Open) & 0.0194 & 0.0447 & 0.1091 \\
12-14 (Market Open) & 0.0193 & 0.0445 & 0.1120 \\
14-16 (Market Open) & 0.0188 & 0.0449 & 0.1065 \\
\bottomrule
\end{tabular}
\end{table}

\section{Conclusion}
Prediction of the stock market using DL has become a popular topic, especially during recent years. If we consider the prediction methods of this rich literature classified into linear and non-linear methods, lack of interpretability and therefore trust in non-linear deep learning-based methods has been facing criticism. In addition, the recent studies of financial markets that are taking advantage of recently developed interpretation techniques, have focused on a very detailed and low-level perspective, such as the importance of words sentiment polarity in tweets regarding stock market.
Instead, our study is devoted to a more high-level interpretation analysis. In this regard, we implemented a DL model using an extensive feature set from two different sources that hold a relatively large number of market's driving factors. Since our proposed feature matrix had a diverse set of features, implementing the explanation method led to a rather comprehensive analysis of the features that are in charge of the movements of the market. Such kind of analysis is prevalent when using a linear method such as the study of \cite{bollen_2011_twitter} \cite{tabari-etal-2018-causality} \cite{abraham_2018_cryptocurrency} which attempted to find out which factor is causing the market movements amongst a set of features. But, this study performed this task within a non-linear framework to benefit from both, deep learning's precision and today's novel interpretation methods. Most importantly, this study is discriminated from the mentioned studies by the fact that each feature in this study is measured by its importance within a bunch of other features, but those studies were analyzing each feature one by one without any other features interfering.

Although this study had innovations in proposing a novel feature matrix, a competitive accuracy in predictions and interpretation of the market's driving factors, it has certain limitations. Firstly, there is a lack of a significance test that can statistically test the importance of the features in our feature set. Secondly, the volume of tweets related to the cryptocurrency market seems to be considerably more than the volume of tweets related to stock market; therefore, conducting the same study on that market may be much more effective. In this respect, we recommend that scholars dedicate their works with the same methodology on popular digital coins. Besides that, we also recommend for future works to work on other companies in the stock market from the tech industry that we feel there are more conversations devoted to them on social media. 
Moreover, as mentioned earlier in the Dataset and pre-processing section, there was a deficit of intra-day data of price due to which we were obligated to interpolate our feature matrix with a copy of the prior day's prices. In this respect, we recommend interested scholars to use 2-hour candles of the prior day to possibly achieve more precise results. In regard to the explanation method, we recommend for future works to use the SHAP method and to compare the results with Lime and other methods of interpretation.

\bibliographystyle{unsrt}  
\bibliography{references}

\end{document}